\documentclass[12pt,a4paper]{cibb}

\makeatletter
\providecommand{\@ordinalM}[2]{#1}
\makeatother

\usepackage{subfigure,graphicx}
\usepackage{amsmath,amsfonts,latexsym,amssymb,euscript,xr}
\usepackage{booktabs}
\usepackage[nodayofweek]{datetime}
\usepackage{hyperref}
\usepackage{fmtcount}
\usepackage[english]{datenumber}
\usepackage[absolute]{textpos}

\usepackage[table]{xcolor}
\usepackage{color,colortbl,tabularx}

\usepackage[english]{babel}
\usepackage[protrusion=true,expansion=true]{microtype}
\usepackage{amsmath,amsfonts,amsthm}
\usepackage{pifont}

\definecolor{LightBlue}{rgb}{0.88,0.9,0.9}

\title{\Large $\ $\\ \bf Treatment persistence drives estimator performance in longitudinal causal inference based on observational data: A simulation study}

\author{\large Sergio Gaiotti$^1$, Sara Poletto$^{1}$, Enrico Longato$^{2,1}$, Erica Tavazzi$^1$, and Martina Vettoretti$^{1*}$}

\address{ \footnotesize $\ $\\$^1$ Department of Information Engineering, University of Padova, Padova, Italy. \\

$^2$ Dept. of Surgical, Oncological, and Gastroenterological Sciences, University of Padova,
Padova, Italy. \\
\bigskip
ORCID codes: SG 0009-0002-1369-0541; SP 0009-0004-4967-3051; EL 0000-0001-5940-645X; ET 0000-0001-6188-6413; MV 0000-0002-5020-1818.
\bigskip
\newline
$^*$corresponding author: martina.vettoretti@unipd.it
}

\abstract{\small longitudinal causal inference, time-varying confounding, treatment persistence. \normalsize
\\[17pt]
{\bf Abstract} Longitudinal clinical data are increasingly available, offering opportunities to study treatment effects over time but also raising challenges related to time-varying confounding and evolving treatment decisions. We investigate how longitudinal treatment dynamics affect causal effect estimation when baseline and longitudinal methods target different causal estimands. Using a structural causal model (SCM), we simulate data with time-varying confounding, binary treatment, and an absorbing binary outcome under 9 scenarios combining functional complexity and treatment persistence. We compare three baseline and two longitudinal estimators against Monte Carlo ground-truth risk ratios (RRs) under sustained treatment regimes. Results show that treatment persistence is the main driver of estimator behaviour. High persistence reduces the discrepancy between baseline and sustained-regime estimands, making baseline estimators closer to the sustained-regime ground truth (average relative deviation of baseline IPTW/TMLE decreasing from 61\% under low persistence to 10\% under high persistence), whereas low persistence induces practical positivity challenges and increases the variability of longitudinal estimators, with empirical 95\% interval widths increasing from 0.22 to 0.46 for longitudinal IPTW and from 0.20 to 0.36 for LTMLE when moving from high to low persistence. These findings emphasise that estimator performance should be interpreted jointly with the target intervention and the treatment process generating the observed data.
}

\begin{document}

\renewcommand{\thefootnote}{}
\footnotetext{\small{Article version: \datedate $\;$ h\currenttime  $\;$ CET}}

\thispagestyle{myheadings}
\pagestyle{myheadings}
\markright{\tt Proceedings of CIBB 2026}

\section{Introduction}
\label{sec:SCIENTIFIC-BACKGROUND}
The increasing availability of longitudinal clinical data, such as electronic health records and disease registries, has expanded opportunities to study treatment patterns, disease progression, and comparative effectiveness in real-world settings. However, using such data to estimate causal effects is challenging because treatment decisions evolve over time and may depend on clinical and treatment history.
This gives rise to time-varying confounding affected by prior treatment~\cite{ROBINS19861393}, for which traditional adjusting methods are generally inadequate. Time-varying covariates can therefore create treatment-confounder feedback, requiring adjustment strategies that respect temporal ordering. Consequently, adjustment strategies must account for the temporal ordering of treatments and covariates to avoid bias. G-methods, including the g-formula, inverse probability weighting for marginal structural models, and structural nested models, were developed to address this problem~\cite{ROBINS19861393}. In practice, however, analysts often rely on simplified approaches that use only baseline information and ignore treatment evolution. This raises the question of how much error is introduced by ignoring longitudinal treatment dynamics, especially since baseline and longitudinal methods generally target different causal estimands. The discrepancy between these estimands is not fixed, but depends on the treatment process and on the regime of interest. For sustained regimes, high treatment persistence, meaning a high probability of remaining on the same treatment status over time, makes the initial treatment increasingly informative about the entire treatment trajectory, whereas frequent switching can make the two estimands substantially different.

A related issue is whether the observed data provide sufficient support for the intervention being targeted. In this context, support refers to the presence of observed treatment histories that are compatible with the target intervention. In longitudinal settings, positivity requires that relevant treatment trajectories occur with non-negligible probability~\cite{hernan2021}. This support depends on the data-generating process and the treatment regime under consideration; in finite samples, its practical impact also depends on sample size. Therefore, treatment persistence plays a dual role, affecting both the alignment between baseline and longitudinal estimands and the empirical support for sustained treatment regimes.
In this work, we study these mechanisms in a controlled simulation setting based on a SCM which represents the data-generating process through structural equations linking each variable to its causal parents and exogenous noise terms~\cite{Pearl_2009}. We vary two features of the data-generating process: the functional form of the structural equations, ranging from linear to nonlinear specifications with interactions, and the degree of treatment persistence. Within this framework, we compare baseline and longitudinal estimators, including a naive method, inverse probability of treatment weighting (IPTW)~\cite{Robins_Hernan_Brumback_2000}, targeted maximum likelihood estimation (TMLE)~\cite{van_der_laan_targeted_2006}, and longitudinal TMLE (LTMLE)~\cite{van_der_laan_targeted_2018}.

Our results show that treatment persistence is a primary driver of estimator behaviour, affecting both estimand alignment and support for sustained regimes. These findings highlight the importance of interpreting estimator performance in light of both the target estimand and the data-generating process.

\section{Data and Methods}
\label{sec:DATA-AND-METHODS}

\subsection{\textbf{Data Generating Process and Simulation Scenarios}}
\label{sec:DATA-GENERATING-PROCESS}
We simulate longitudinal data from an SCM defined over discrete time points $t = 0, \dots, T$. The simulation design was chosen to represent a minimal but nontrivial longitudinal setting with baseline confounding, time-varying confounding affected by prior treatment, treatment feedback, and an absorbing binary outcome, meaning that once the outcome switches to 1, it remains 1 thereafter. At baseline, we observe covariates $B = (B^1, B^2, B^3)$. At each time $t$, we observe time-varying covariates $L_t = (L_t^1, L_t^2)$, a binary treatment $A_t$, and a binary outcome $Y_{t+1}$, representing the occurrence of the event of interest after time $t$. The exogenous variables $U^{B}, U_t^L, U_t^A,$ and $U_{t+1}^Y$ represent unexplained variation in the corresponding structural equations and are assumed mutually independent, implying the absence of unmeasured confounding. No censoring is assumed, so all subjects remain under observation until the end of follow-up.

The data-generating process is specified through the following system of structural equations:
    \begin{align}
        B &= f_{B}(U^{B}) \\
        L_t &= f_{L_t}(B, L_{t-1}, A_{t-1}, U_t^L) \\
        A_t &= f_{A_t}(B,L_{t}, \overline{A}_{t-1}, U_t^A) \\ 
        Y_{t+1} &= f_{Y_{t+1}}(B, L_t, \overline{A}_t, U_{t+1}^Y)
    \end{align}
where $\overline{X}_t = (X_0, \dots, X_t)$ denotes the history of variable $X$ up to time $t$. 
This specification encodes the temporal and causal structure of the system: covariates follow a first-order Markov process (conditional on baseline), treatment depends on current covariates and past treatment history, and the outcome depends on current covariates and treatment history. 

We consider a finite horizon $T = 4$ and assume time-invariant structural equations, so that functional forms and parameters do not vary across time. 
Treatment is specified to have a protective effect on the outcome, as is natural in clinical settings where a therapy is intended to reduce event risk. Additionally, we calibrate the data-generating process so that the marginal probability of treatment increases from 0.3 at baseline to 0.5 at the end of follow-up, reflecting settings in which treatment prevalence increases over time, as may occur after the introduction or broader adoption of a therapeutic option.

We use a parametric specification that separates dependence structure, nonlinearity, and interaction effects.
Specifically, for each node $X_j$ of the SCM, with causal parents $X_{\mathrm{pa}(j)}$, we assume a parametric structural equation of the form
    \begin{equation}
    \mathbb{E}[X_j \mid X_{\mathrm{pa}(j)}] = g_j\big(h_j(X_{\mathrm{pa}(j)})\big),
    \end{equation}
where the function $h_j : \mathbb{R}^{|X_{\mathrm{pa}(j)}|} \to \mathbb{R}$ defines a scalar predictor based on the parent variables, and $g_j : \mathbb{R} \to \mathbb{R}$ is a node-specific link function, defined as the identity for continuous variables and the inverse-logit for binary variables. For continuous variables, structural equations follow an additive Gaussian noise model. For binary variables, randomness is introduced through the conditional Bernoulli distribution, with no explicit additive noise term on the observed scale.
The predictor $h_j$ is further decomposed as
    \begin{equation}
    h_j(X_{\mathrm{pa}(j)}) = h_j^{(1)}\big(\eta_j(X_{\mathrm{pa}(j)})\big)
    \ , 
    \quad
    \eta_j(X_{\mathrm{pa}(j)}) = w_j^\top T_j(X_{\mathrm{pa}(j)}).
    \end{equation}
This decomposition separates three key components of the data-generating process:
(i) the transformation $T_j$ of the parent variables, which can introduce nonlinear features and interactions;
(ii) the coefficients $w_j$, which control the strength and direction of dependencies;
(iii) the function $h_j^{(1)}$, which allows for additional nonlinear transformations of the linear predictor $\eta_j$.

Within this framework, we obtain nine simulation scenarios by considering two axes of variation. The first axis controls the functional specification of the structural equations through $T_j$ and $h_j^{(1)}$, thereby modulating nonlinearity and interaction effects. The second axis controls treatment persistence through the treatment-history coefficients $w^A_{k,t}$.

In the linear scenario, both $T_j$ and $h_j^{(1)}$ are identity functions. In the nonlinear scenario, nonlinearities are introduced through $T_j$, while $h_j^{(1)}$ remains the identity. In the nonlinear with interactions scenario, $T_j$ additionally includes interaction terms, and the outcome equation allows higher-order nonlinearities through $h_j^{(1)}$.

With respect to the second axis, to control treatment persistence we model the dependence of treatment assignment at time $t$ on its past history $\overline{A}_{t-1}$ through a distributed lag structure in the linear predictor of the treatment model. 
For $t \geq 1$, let $\eta_t^A$ denote the linear predictor for $A_t$. We write
\begin{equation}
    \eta_t^A
    =
    r_t^A(B,L_t)
    +
    \sum_{k=0}^{t-1} w_{k,t}^A A_k,
    \qquad
    w_{k,t}^A = w_{t-1,t}^A b^{t-1-k}.
\end{equation}
where, $r_t^A(B,L_t)$ captures the contribution of baseline covariates and current time-varying covariates to treatment assignment, $w_{t-1,t}^A$ controls dependence on the most recent treatment, and $b \in [0,1]$ governs the decay of past treatment influences. Larger values of $w_{t-1,t}^A$ and $b$ induce higher persistence in treatment trajectories, making treatment assignment increasingly determined by its past values.
To define distinct persistence regimes, we jointly vary these parameters and calibrate them to achieve target levels of switching in the observed data. Specifically, we consider the marginal probability that a subject experiences at least one treatment switch during follow-up:
    \begin{equation} p_{\mathrm{sw}}
    =
    \Pr\left(\sum_{t=1}^{T} \mathbb{I}(A_t \neq A_{t-1}) > 0\right), \end{equation}
where $\mathbb{I}$ denotes the indicator function,
and set it to approximately 0.7, 0.45, 0.25, corresponding to the low-, medium-, and high-persistence scenarios, respectively.

\subsection{\textbf{Ground Truth}}
\label{sec:GROUND-TRUTH}
We consider the effect of sustained treatment on the risk of experiencing the outcome by the end of follow-up. We define two static treatment regimes:
    \begin{align}
        g_0 &: A_t = 0 \quad \forall t \in \{0,\dots,T\}, \\
        g_1 &: A_t = 1 \quad \forall t \in \{0,\dots,T\},
    \end{align}
corresponding to the ``never treat'' and ``always treat'' strategies, respectively.

Let $Y^{g}$ denote the counterfactual outcome at time $T+1$ under regime $g$. The causal estimand of interest is the risk ratio (RR):
    \begin{equation}
       \text{RR} = \frac{\mathbb{E}[Y^{g_1}]}{\mathbb{E}[Y^{g_0}]}. \label{eq:RR}
    \end{equation}

We approximate these expectations by Monte Carlo under each treatment regime. For $g \in \{g_0,g_1\}$, treatment mechanisms are deterministically set according to $g$, while all other structural equations are left unchanged, and
\begin{equation}
\mathbb{E}[Y^g] \approx \frac{1}{N}\sum_{i=1}^{N} Y_i^{(g)},
\end{equation}
where $Y_i^{(g)}$ denotes the simulated counterfactual outcome for unit $i$ under regime $g$.
The resulting ground-truth RR is used as a benchmark for the estimators described below.

\subsection{\textbf{Estimands and Estimators}}

We consider two classes of estimators that differ in the causal estimand they target. 
Baseline estimators rely exclusively on information available at time $t=0$ and target the causal effect of a one-time intervention on the initial treatment $A_0$. Specifically, they aim to estimate the risk ratio comparing the counterfactual outcomes under $A_0 = 1$ versus $A_0 = 0$, without controlling for subsequent treatment decisions:
    \begin{equation} \frac{\mathbb{E}[Y^{A_0=1}]}{\mathbb{E}[Y^{A_0=0}]} \end{equation} 
We also include a naive unadjusted estimator, which compares observed mean outcomes between individuals with $A_0=1$ and $A_0=0$, as an associational benchmark.
We consider two methods: a baseline-only IPTW, based on the propensity score $P(A_0 \mid B)$, and baseline TMLE,  a doubly robust estimator combining an outcome regression with the same propensity score model.

In contrast, longitudinal estimators account for the full treatment history and are aligned with the sustained intervention regimes defined in Section~\ref{sec:GROUND-TRUTH}.
Longitudinal IPTW constructs weights based on the product of conditional treatment probabilities across time, to adjust for time-varying confounding affected by prior treatment. LTMLE combines sequential outcome regression and treatment modelling in a doubly robust framework.

We use the \texttt{ltmle} R package~\cite{ltml_r_package}. Baseline and longitudinal IPTW estimate the corresponding treatment mechanisms using logistic regression. Baseline TMLE and LTMLE estimate outcome regressions and treatment mechanisms using Super Learner~\cite{vanderLaanPolleyHubbard+}, with library \texttt{SL.glm}, \texttt{SL.stepAIC}, \texttt{SL.nnet}, \texttt{SL.gam}, and \texttt{SL.glmnet}. For each scenario, we simulate 100 datasets of $n=1000$ individuals and compare RR estimates with ground-truth values approximated from an independent Monte Carlo sample of size $N=10^6$.

\section{Results and Discussion}
\label{sec:RESULTS}

\begin{figure}[h]
    \vspace{3mm}
     \begin{center}
     \includegraphics[width=.9\textwidth, trim=20 7 12 13, clip]{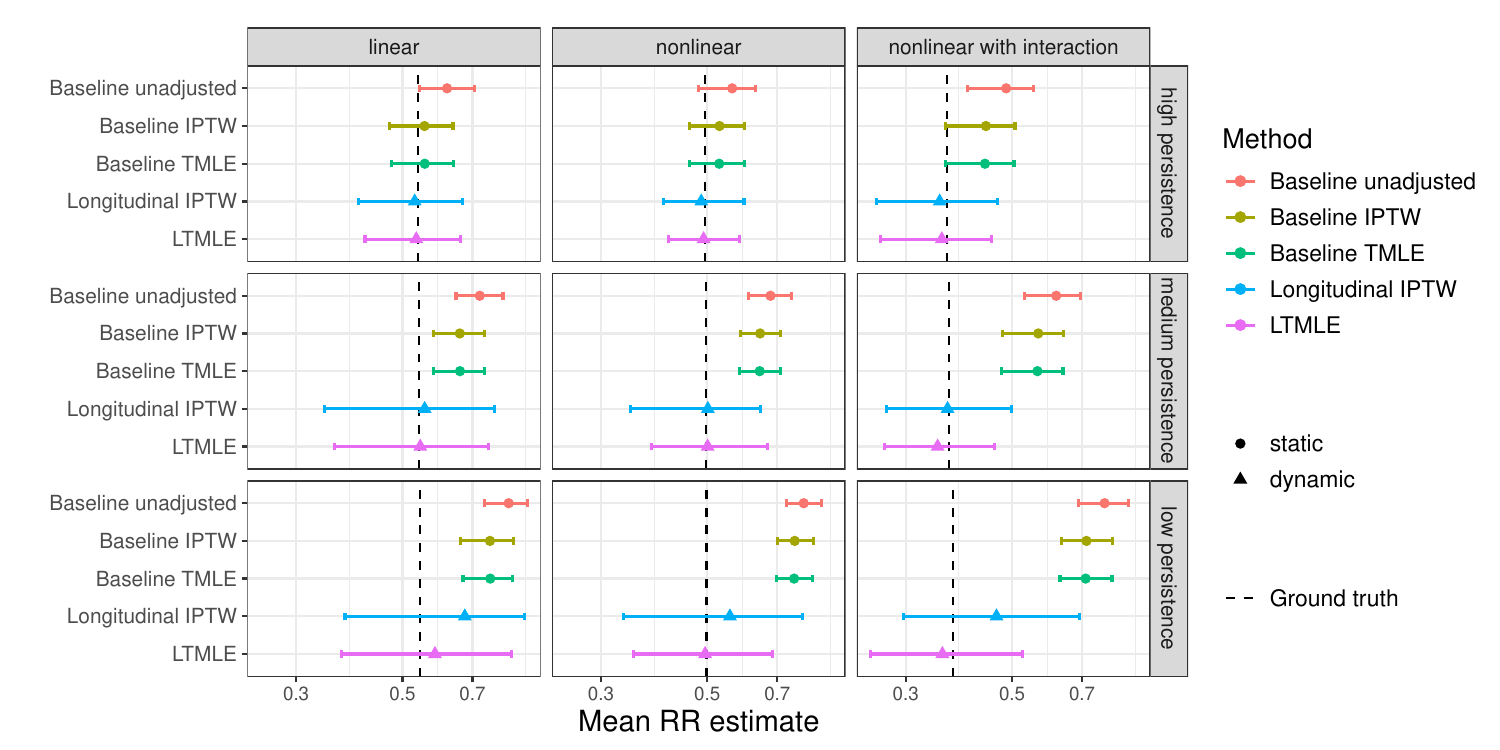}
    \caption{Mean estimated RR and empirical 95\% simulation intervals. Columns correspond to functional forms and rows to treatment-persistence levels. Points denote mean estimates over 100 simulated datasets; error bars indicate empirical 2.5th and 97.5th percentiles across replicates. The dashed vertical line indicates the ground-truth RR. The x-axis is on a log$_{10}$ scale.}
    \label{fig:RR}
    \end{center}
    \vspace{-8mm}
    \end{figure}

Figure \ref{fig:RR} summarises the mean estimated RR and empirical 95\% simulation intervals across the nine simulation scenarios. 

A dominant pattern emerges along the treatment persistence axis. For baseline IPTW and baseline TMLE, the average relative deviation from the sustained-regime ground-truth RR, computed as the difference between the estimate and the sustained-regime ground truth divided by the ground-truth RR, decreases from approximately 61\% in the low-persistence setting to 35\% and 10\% in the medium- and high-persistence settings, respectively.  For the unadjusted baseline estimator, the corresponding values are approximately 73\%, 46\%, and 21\%. This reflects a convergence between the causal estimand targeted by baseline methods and the estimand under sustained treatment regimes. Indeed, when treatment is highly persistent, the initial assignment $A_0$ becomes highly predictive of the full treatment trajectory $\overline{A}_T$, so that interventions on $A_0$ approximate sustained interventions on the entire treatment path.

Longitudinal estimators, which target the sustained regimes, exhibit substantial variability in low-persistence scenarios across all functional forms. This variability decreases as persistence increases.
Averaged across functional-form scenarios, the 95\% Monte Carlo interval width of longitudinal IPTW estimates decreases from 0.46 under low persistence to 0.22 under high persistence, while for LTMLE it decreases from 0.36 to 0.20. This pattern is consistent with practical positivity limitations: under frequent switching, sustained trajectories are rare, reducing empirical support and increasing estimator instability.

In low-persistence settings, longitudinal IPTW exhibits upward bias across all functional forms, with relative bias ranging from approximately 12\% to 24\%, whereas LTMLE remains closer to the ground truth, with relative bias ranging from approximately -5\% to 8\%. This behaviour may be attributable to finite-sample bias induced by highly variable weights and, potentially, to misspecification of the treatment model. This effect appears to be mitigated in LTMLE, likely due to its doubly robust construction.

Differences across functional forms are comparatively modest. Increasing nonlinearity appears to slightly degrade performance, particularly for baseline estimators, although the overall behaviour is primarily driven by treatment persistence.

\section{Conclusion}
\label{sec:CONCLUSIONS}
Our results show that treatment persistence plays a dual role: it influences both the discrepancy between baseline and sustained-regime estimands and the empirical support for sustained treatment trajectories. In high-persistence settings, baseline estimands increasingly approximate sustained-regime estimands, reducing the bias of baseline methods. In low-persistence settings, this alignment breaks down, while longitudinal estimators may exhibit increased variability due to limited support.
Several design choices were made to keep the simulation setting controlled and interpretable. In particular, we considered settings without censoring, used a fully parametric data-generating process, and focused on a short time horizon. Future work could extend this framework to longer follow-up, censoring, dynamic treatment regimes, stronger positivity challenges, and methods that treat deviations from a target regime as censoring.

\section*{Conflict of interests}
\label{sec:CONFLICT-OF-INTERESTS}
Nothing to declare.


\section*{Funding}
\label{sec:FUNDING}
This work was funded by the REDDIE (Real-world evidence for decisions in diabetes) project. The REDDIE project has received funding from the European Union’s Horizon 2022 research and innovation programme under grant agreement No. 101095556. Views and opinions expressed are however those of the author(s) only and do not necessarily reflect those of the European Union or European Health and Digital Executive Agency (HADEA). Neither the European Union nor the granting authority can be held responsible for them. This work has received funding from the UK research and Innovation under contract number 101095556.


\footnotesize
\bibliographystyle{unsrt}
\bibliography{bibliography_CIBB_file.bib} 
\normalsize

\end{document}